\documentclass[useAMS, usenatbib] {mn2e}
\usepackage{graphicx}
\usepackage{longtable}
\usepackage{lscape}
\usepackage{pdfsync}
\usepackage{bbding}
\usepackage{natbib} 
\usepackage{txfonts}
\DeclareGraphicsExtensions{.pdf,.png,.jpg}

\title[Interacting LAEs at z = 5.1]{\bf {Episodic star formation in a group of LAES at z= 5.07} }
\author [J. M. Rodr\'iguez Espinosa et al.] {J.M.~Rodr\'iguez Espinosa,$^{1,2}$\thanks{E-mail: jmr.espinosa@iac.es}
O.~Gonz\'alez-Mart\'in,$^{1,2}$\thanks{Juan de la Cierva Fellow} N.~Castro Rodr\'iguez,$^{1,2}$ 
\newauthor
P. G. P\'erez-Gonz\'alez,$^3$ J.M. Mas-Hesse,$^4$ C.~Mu\~noz-Tu\~n\'on,$^{1,2}$ A. Cava,$^5$  N. Cardiel,$^3$ 
\newauthor
A. Cabrera Lavers,$^6$ J. Gallego,$^3$ A. Hern\'an Caballero,$^7$ N. Herrera Ruiz,$^8$
\newauthor
and N. Ram\'irez Olivencia$^3$ \\ 
$^1$Instituto de Astrof\'isica de Canarias, E-38200 La Laguna, Spain \\ 
$^2$Departamento de Astrof\'sica, Universidad de La Laguna, E-38205 La Laguna, Spain\\
$^3$Departamento de Astrof\'isica y CC de la Atm\'osfera, Universidad Complutense de Madrid, Spain\\
$^4$Centro de Astrobiolog\'ia - Dept. de Astof\'isica (CSIC-INTA), Madrid, Spain\\ 
$^5$Observatoire de Gen{\`e}ve, Universit{\'e} de Gen{\`e}ve, 51 Ch. des Maillettes, 1290 Versoix, Switzerland\\
$^6$GRANTECAN, 38712 La Palma, Spain \\
$^7$ Instituto de F\'isica de Cantabria (IFCA), CSIC-UC, E-39005 Santander, Spain \\
$^8$Astronomisches Institut der Universit\"at Bochum, Universit\"atsstr. 150, 44801 Bochum, Germany}

\begin{document}

\date{submitted 2014 ...}

\pagerange{\pageref{firstpage}--\pageref{lastpage}} \pubyear{...}

\newtheorem{theorem}{Theorem}[section]

\maketitle

\label{firstpage}

\begin{abstract} 
We are undertaking a search for high-redshift low luminosity Lyman Alpha sources in the SHARDS survey. Among the pre-selected Lyman Alpha sources 2 candidates were spotted, located 3.19'' apart, and tentatively at the same redshift. Here we report on the spectroscopic confirmation with GTC of the Lyman Alpha emission from this pair of galaxies at a confirmed spectroscopic redshifts of z$=$5.07. Furthermore, one of the sources is interacting/merging with another close companion that looks distorted. Based on the analysis of the spectroscopy and additional photometric data, we infer that most of the stellar mass of these objects was assembled in a burst of star formation 100~Myr ago. A more recent burst (2~Myr old) is necessary to account for the measured Lyman Alpha flux. We claim that these two galaxies are good examples of Lyman Alpha sources undergoing episodic star formation. Besides, these sources very likely constitute a group of interacting Lyman Alpha emitters (LAEs).

\end{abstract}


\section{Introduction}

The study of high redshift galaxies is important for a better understanding of the formation of structure in the Universe, and of the processes conducting to the complete re-ionisation of the intergalactic neutral hydrogen pervading the early Universe. The past decades have seen rapid progress in the investigation of high-z galaxies, often due to better instrumentation in large telescopes, and the implementation of novel observing techniques \citep{Bromm&Yoshida2011}. These are principally broad-band colour selection criteria, and the use of narrow passband filters (\citealt{Steideletal1999}; \citealt{Steideletal2003}; \citealt{Giavaliscoetal2004}; \citealt{Taniguchietal2005}; \citealt{Stark2009}; \citealt{Bouwens2009}; \citealt{Bouwens2010}; \citealt{Ouchi2010}). Recently, with the new {\it {HST}} camera WFC3, the redshift range has been extended further, and new high redshift galaxies at $6.5 \leq z \leq 10$  have been detected (e.g., \citealt{Bunker2010}; \citealt{McLure2010}; \citealt{Oesch2010}; \citealt{Yan2011}; \citealt{Bouwens2011}; \citealt{Bouwens2013}; \citealt{Ellis2013}; \citealt{Laporte2014}). Furthermore, although challenging, spectroscopic confirmation of high-redshift Ly$\rm{\alpha}$ sources is now achievable with efficient spectrographs in large telescopes (e.g, \citealt{Yuma2010}; \citealt{Stark2010}; \citealt{Stark2011}; \citealt{Hu2010}; \citealt{Kashikawa2011}; \citealt{Curtis-Lake2012};  \citealt{Curtis-Lake2013}). 

Another aspect to probe in high redshift sources is their clustering properties. In the $\Lambda$CDM paradigm, galaxy interactions and mergers play a key role in galaxy evolution. Not much literature is available focusing on the clustering  properties and interactions at z$>$5. However, the visibility of Ly$\rm{\alpha}$ sources at very high redshifts is boosted if these sources were clustered (\citealt{Miralda-Escude1998}; \citealt{McQuinn2007}). Note that clustering also influences the possible modes of star formation, favouring short star formation timescales \citep{Lee2009}. 

The recent Survey for High-z Absorption Red and Dead Sources (SHARDS\footnote[1]{http://guaix.fis.ucm.es/$\sim$pgperez/SHARDS/}) \citep{Perez-Gonzalez2013}, opens an alternative and powerful tool to select high-z candidates, thanks to its depth and spectral resolution. SHARDS allows to confirm the presence of emission lines (thus their redshift), at very faint magnitudes (down to 27 mag AB from 500 to 950 nm), even beyond the spectroscopic limit. Here we present a pair of sources at z$\sim$5.1 selected with SHARDS. Their relevance is based on the fact that {\bf they seem to be} an interacting group, and that they are undertaking episodic star formation bursts.Ê Section 2 briefly describes the SHARDS project and our survey for very high-z Ly$\rm{\alpha}$ sources. Section 3 describes the GTC spectroscopy. Section 4 discusses the main properties of the two z$\sim${}5 objects analyzed and Section 5 summarizesÊ the main results. We adopt a $\Lambda$-dominated, flat universe, with $\Omega_\Lambda = 0.73$, $\Omega_M =0.23$, and $H_{\rm 0} = 71$ km s$^{-1}$ Mpc$^{-1}$. Magnitudes are given in the AB system \citep{Oke&Gunn1983}.
 
\section{The SHARDS survey}

SHARDS is a medium-band optical survey carried out with the Spanish 10.4m GTC. The OSIRIS instrument was used equipped with 24 custom made contiguous medium-band filters (typical width 17 nm) spanning the wavelength range between 500 and 950 nm, with an effective spectral resolution R$\gtrsim$50. A detailed discussion of the SHARDS survey, the data reduction and calibration procedures can be found in \citet{Perez-Gonzalez2013}. SHARDS covered the entire GOODS North field ($\sim$~141~arcmin$^2$) in two pointings, including most of the area observed with the {\it {HST/ACS}} and {\it WFC3} instruments by the GOODS and CANDELS surveys. 

\begin{table}
\caption{SHARDS plus additional photometric data for the sources, SHARDS123744.98+621820.2 (Obj1) and SHARDS123744.84+621817.2 (Obj2). Magnitudes are in the AB system. Wavelengths are in nm. Filters starting with S are SHARDS filters. {\bf Upper limits are 1$\sigma$.}}
\label{Tab1}
\begin{tabular}{@{}lcccc}

Filter & CWL & FWHM & Obj1 & Obj2 \\
\hline
SF619W17 & 618.7 & 15.8 & $<26.9$ & $<26.9$ \\
SF636W17 & 636.4 & 16.2 & $<26.8$ & $<26.8$ \\
SF653W17 & 653.4 & 15.4 & $<26.9$ & $<26.9$ \\
SF670W17 & 670.4 & 16.0 & $<26.6$ & $< 26.6$ \\
SF687W17 & 686.9 & 17.2 & $<27.0$ & $<27.0$ \\
SF704W17 & 703.7 & 17.9 & $<26.7$ & $<26.7$ \\
SF721W17 & 723.1 & 18.5 & $26.37 \pm 0.49$  &  $<26.4$  \\
SF738W17 & 740.7 & 14.9 & 24.84 $\pm 0.06$ & 25.36 $\pm 0.11$ \\ 
SF755W17 & 757.5 & 15.3 &  25.85 $\pm 0.23$ & 26.08 $\pm 0.21$ \\
SF772W17 & 767.7 & 15.8 &  $<26.5$ & $25.99 \pm 0.18$ \\
SF789W17 & 790.7 & 16.0 & 25.90 $\pm 0.26$ & $26.10 \pm 0.30$  \\
SF806W17 & 802.5 & 16.1 & $<26.4$ & 26.09 $\pm 0.19$ \\
SF823W17 & 828.1 & 14.7 &  $25.60 \pm 0.23 $ & $<26.7 $ \\
SF840w17 & 842.1 & 15.6 & $26.00 \pm 0.29$ & 26.05 $\pm 0.40$ \\
SF857W17 &852.4 & 15.9 & $<26.6 $ & 26.07 $\pm 0.13$\\
SF883W35 & 883.8 & 33.6 & $ 25.86 \pm 0.31$ & 25.90 $\pm 0.28$ \\
SF941W33 & 942.4 & 33.3 & $ < 26.2 $ &  26.10 $\pm 0.59$ \\
ACSF775W & 770.6 & 132.2 & 25.89 $\pm 0.13$ & 26.01 $\pm 0.07$ \\
ACSF850LP & 905.3 & 126.5 & 26.12 $\pm 0.23$ & 26.01 $\pm 0.07$ \\
WFC3F105W & 1058.4 & 265.2 &  25.80 $\pm 0.24$ & 25.98 $\pm 0.17$ \\
WFC3F125W & 1251.6 & 284.5 &  26.80 $\pm 0.15$ & 25.97 $\pm 0.15$ \\
WFC3F160W & 1539.2 & 268.2 & 26.09 $\pm 0.18$ & 26.03 $\pm 0.14$ \\
IRAC36 & 3557.2 & 683.8 & 26.00 $\pm 0.25$ & 26.10 $\pm 0.19$ \\
IRAC45 & 4504.9 & 865.1 & 26.20 $\pm 0.30$ & 26.20 $\pm 0.20$ \\
 \end{tabular}
 
  \end{table}

\begin{table*}
\caption{Main parameters derived from the GTC spectroscopy. The M$_{UV}$ and rest frame EW have been derived from the SHARDS photometric data, as the spectroscopic continuum has not been detected. {\bf The SFRs have been computed from the Ly$\alpha$ fluxes, using \citet{Kennicutt1998}, to facilitate their comparison with other results in the literature.}}
\label{Tab2}
\begin{tabular}{@{}lcccccccc}
 
 & R.A.(J2000) & Dec.(J2000) & Redshift & Ly$\rm{\alpha}$ flux &  L$_{\rm{Ly\rm{\alpha}}}$ & SFR & M$_{UV}$ & EW$_{rf}$ \\
 & & & &  $erg s^{-1} cm^{-2}$  & $erg/s$ & $M_{\sun}/yr$ & & $\AA$ \\
 \hline
 Obj1 & 12:37:45.02 & +62:18:20.33 & $5.0722 \pm 0.0012$ & $1.63 \pm 0.29 \times 10^{-17}$ & $4.57 \pm 0.81\times 10^{42}$ & $4.2 \pm0.7$  & -22.4 $\pm 0.4$ & $41\pm 5$ \\
 Obj2 & 12:37:44.87 & +62:18:17.30 & $5.0754 \pm 0.0012$ & $6.85 \pm 1.74 \times 10^{-18}$  & $1.92 \pm 0.49\times 10^{42}$ & $ 1.8 \pm 0.5 $ & -22.8 $\pm 0.6$ & $26\pm 5$\\
 \end{tabular}
 \end{table*}
 
We have benefited both from the depth and spectral resolution of the SHARDS survey to look for both Lyman Alpha Emitters (LAEs) and Lyman Break Galaxies (LBGs). Our technique for looking for Ly${\rm{\alpha}}$ sources has used the possibilities offered by the SHARDS set of contiguous filters to search for drop out sources and objects appearing in just one filter. The process included visual inspection of the candidates to sort out artifacts or cosmetic defects in the arrays. Further details about the procedures and results of the survey for very high-z sources will be presented in Rodr\'iguez Espinosa et al. (2014), in preparation.

 \begin{figure} 
	\includegraphics[width=28mm]{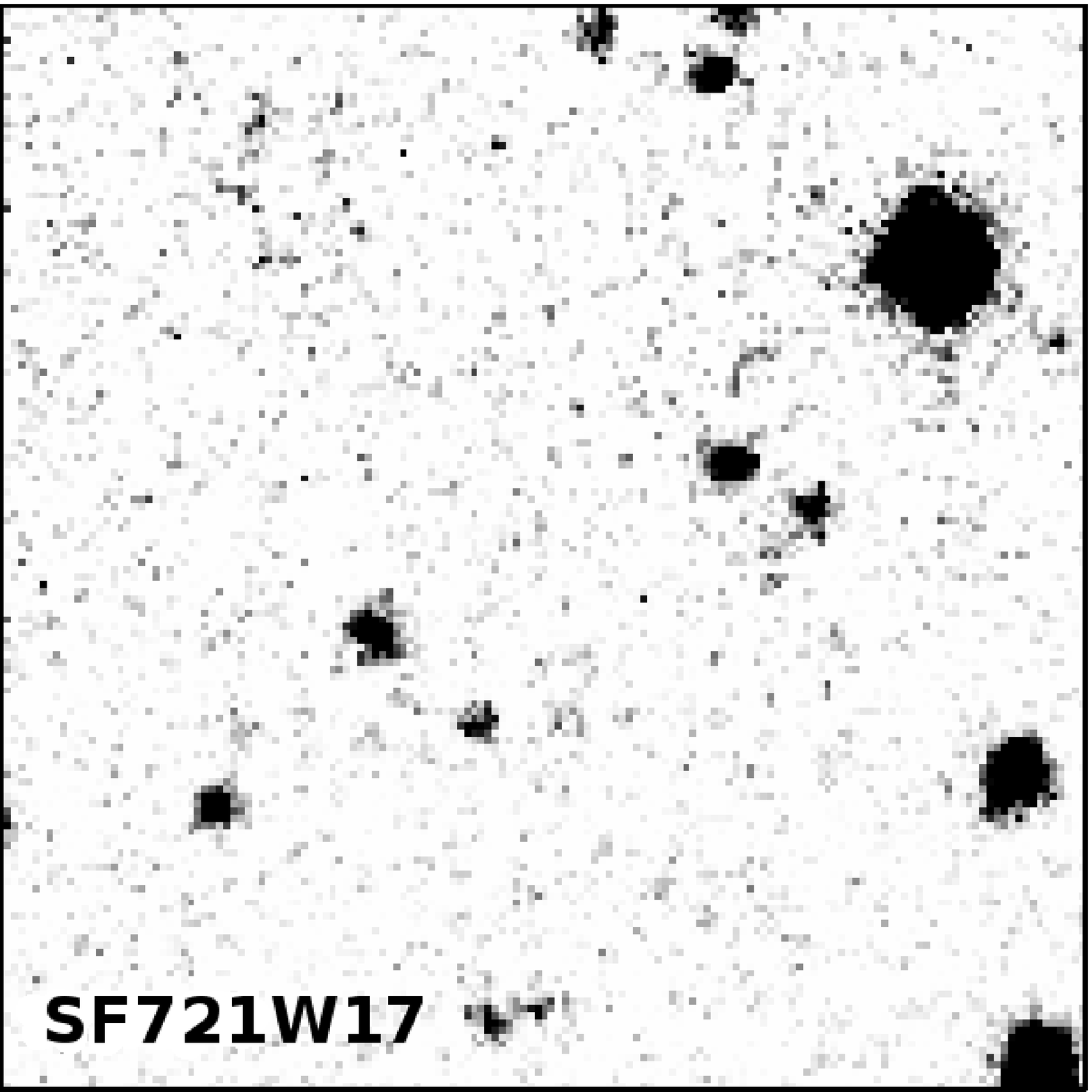}\includegraphics[width=28mm]{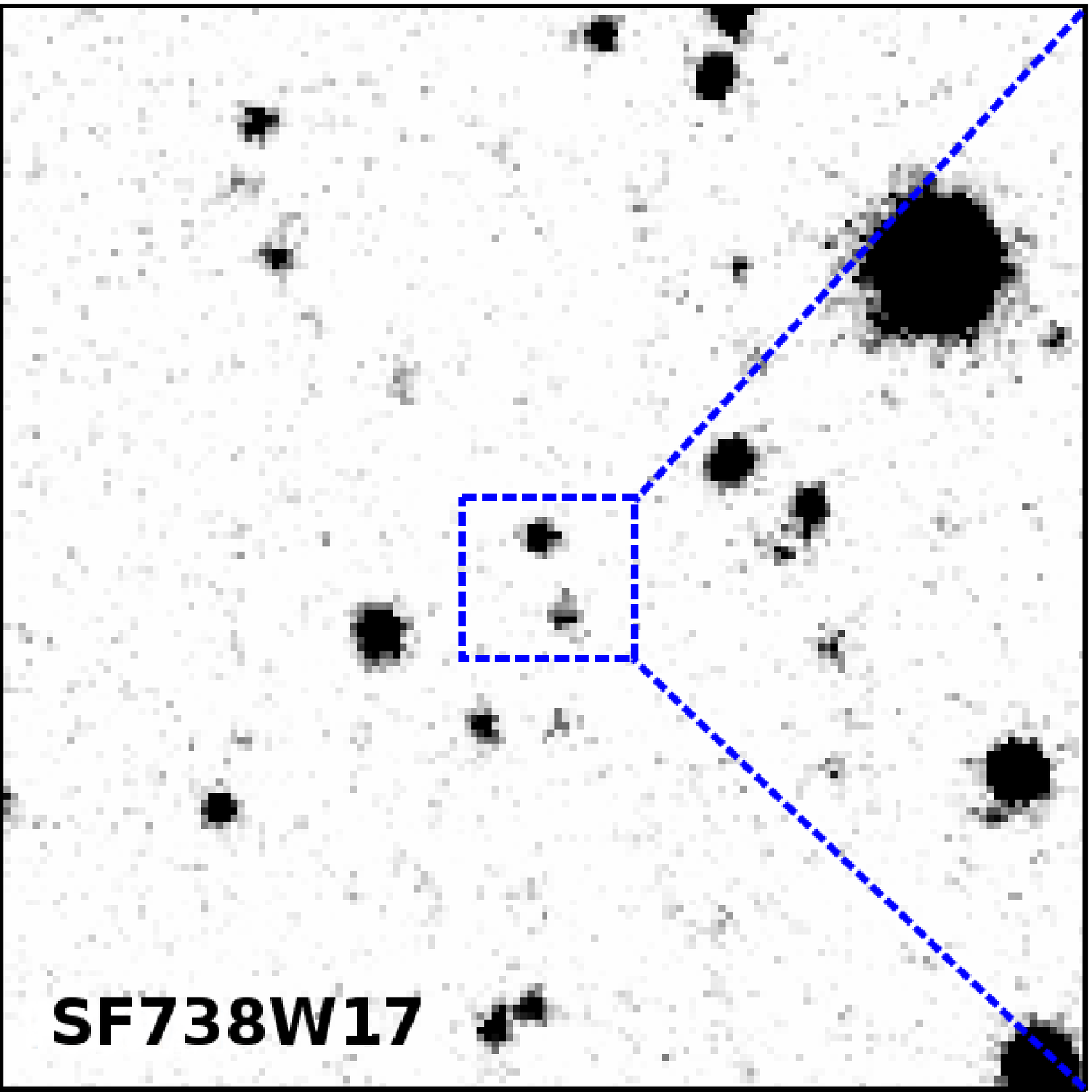}\includegraphics[width=28mm]{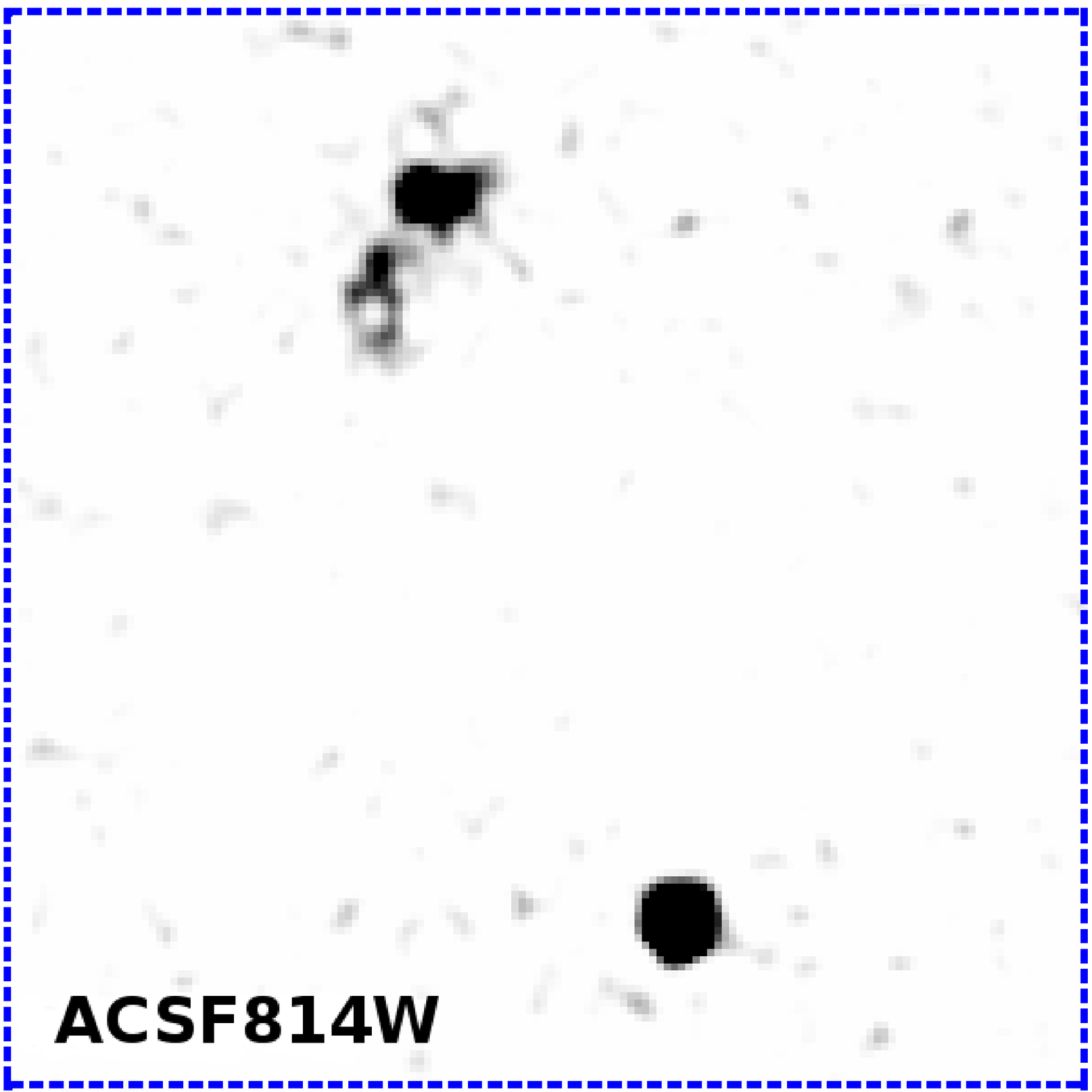} 
	\caption{SHARDS images of the two candidate sources in filters SF721W17 (left) and SF738W17 (center). Note the two conspicuous sources in the latter filter not visible in the bluer one.The right panel shows an 814~nm {\it HST/ACS} zoomed image, smoothed and contrast improved, to show the nebulosity around the northern source (Obj1). North is up, east is left.} 
	\label{Twosources} 
\end{figure}

Among the candidates selected, we identified two sources, separated by only 3.19 arcsec, that appeared prominently in the SF738W17 filter, but not in any other bluer filter (Fig.~\ref{Twosources}). Both sources were however detected in several SHARDS filters red-wards of SF738W17, as well as in {\it HST/ACS} deep images, in three {\it HST/WFC3} bands, and in two {\it Spitzer/IRAC} bands (see Tab.~\ref{Tab1}). Furthermore, the high resolution {\it HST} images reveal an extended structure in the northern object, pointing towards the position of the southern galaxy (see Fig.~\ref{Twosources} right panel). This distorted configuration seems to be either a tidal tail associated to the interaction with the southern object, or rather the result of a recent merger with a third source. Although these sources were detected primarily for their prominent emission in the SF738W17 filter, their identifications in additional filters to the red was crucial in showing their rest frame UV -Optical continua typical of LBGs. We have thus found two seemingly interacting LBGs that show Ly${\rm{\alpha}}$ emission. These sources carry the SHARDS catalogue names, SHARDS123744.98+621820.2 and SHARDS123744.84+621817.2, though in what follows, these sources will be referred to as Obj1 and Obj2, respectively.

\section{GTC spectroscopy}

\begin{figure} 
	\begin{centering}
	\includegraphics[width=85mm]{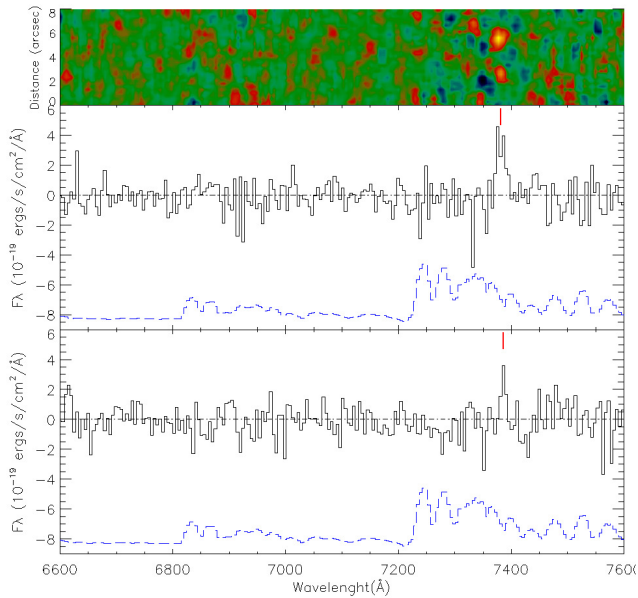} 
	\caption{Upper Panel: 2-D spectral image. Center Panel: Obj1. Lower Panel: Obj2. The Ly$\rm{\alpha}$ lines, marked with a red line, are clearly seen in both spectra. The blue dashed line, at the bottom of each frame, shows an off-scale sky spectrum. Notice that the continuum has not been detected in any of the spectra.} 
	\label{Twospectra} 
	\end{centering}
\end{figure}

These two objects were perfect candidates for long-slit spectroscopy with the GTC, in absence of the multi-object capability not available at the time of the observations. The spectra were therefore taken with the slit passing through both objects. The observations were acquired in February 7, 2013, using OSIRIS at the GTC. We used the R500R grating, producing a dispersion of 4.88 \AA$/$pix, and covering the wavelength range from 500 to 970 nm.  Integration time was 6400 second, distributed in 4 dithered 1600 sec frames. Seeing was 0.8 arcsec. The slit width employed was 1.21 arcsec.  We used the standard star Ross 640 for flux calibration, and suitable arc lines for wavelength calibration. Data reduction followed the usual procedures using standard IRAF tasks. The most difficult part was the sky subtraction, as the region between 680-1000 nm is full of sky lines. We knew from the SHARDS photometry that both sources should pick strongly around 738 nm. So we centered our efforts around this area in the two-dimensional image (see Fig.~\ref{Twospectra}, top panel). To get rid of the sky contamination in this area, we subtracted iteratively the sky using sky regions adjacent to the object positions. Fig.~\ref{Twospectra} shows the extracted 1-D spectra for both objects. A clear emission line (S/N of 6  \& 4, respectively) is seen in each of the spectra within the wavelength range of the F738W17 filter.  To recover the line fluxes we have accounted for light losses (a factor of 1.23)  in the slit. We assume that the emission lines shown in Fig.~\ref{Twospectra} must be Ly$\rm{\alpha}$, based on the SEDs of the candidate sources that clearly show a strong drop identified as the Lyman break. Moreover, no sign of continuum emission is seen in deep broad band images blue-wards of the Ly$\rm{\alpha}$ line, while the sources can be readily seen in many other redder filters.   

\section{Results}

Table~\ref{Tab2}  gives positions,  fluxes, luminosities and redshifts for both sources. The best values obtained for the redshifts are $z= 5.0722 \pm 0.0012$ and $z= 5.0754 \pm 0.0012$ for Obj1 and Obj2 respectively. At this redshift, the rest-frame UV absolute magnitudes are M$_{UV} (\simeq$ M$_{\rm SF755W17}$) = (-22.4) for Obj1, and (-22.8) for Obj2. Both sources are close in velocity space ($\sim$159 km/s), with a  separation in the sky of only $3.19$ arcsec, corresponding to a physical separation of 20.3 kpc (the scale at $z\sim$~5 is 6.37 kpc/arsec). We therefore conclude that these two sources form a close pair at $z\sim$~5. Furthermore, as mentioned before, Obj1 seems to be interacting or  merging with yet another galaxy, located at less than 3 kpc to the SE from its nucleus, as can be seen in the {\it HST} image (right panel in Fig.~\ref{Twosources}).  Although we claim below that the star formation in these objects is episodic and short-lived, we include in Table~\ref{Tab2} {\bf a derived star formation rate (SFR)} following \citet{Kennicutt1998}, for comparison with other results in the literature, {\bf where typically no distinction is made between constant star formation or episodic star formation}. The derived SFRs are in agreement with observations of similar {\bf (in terms of UV luminosity and rest-frame EW)} high z sources, ({\it e.g.} \citealt{Cassata2011}). 

\begin{figure} 
	\begin{centering}
	\includegraphics[angle=-90, width=\columnwidth]{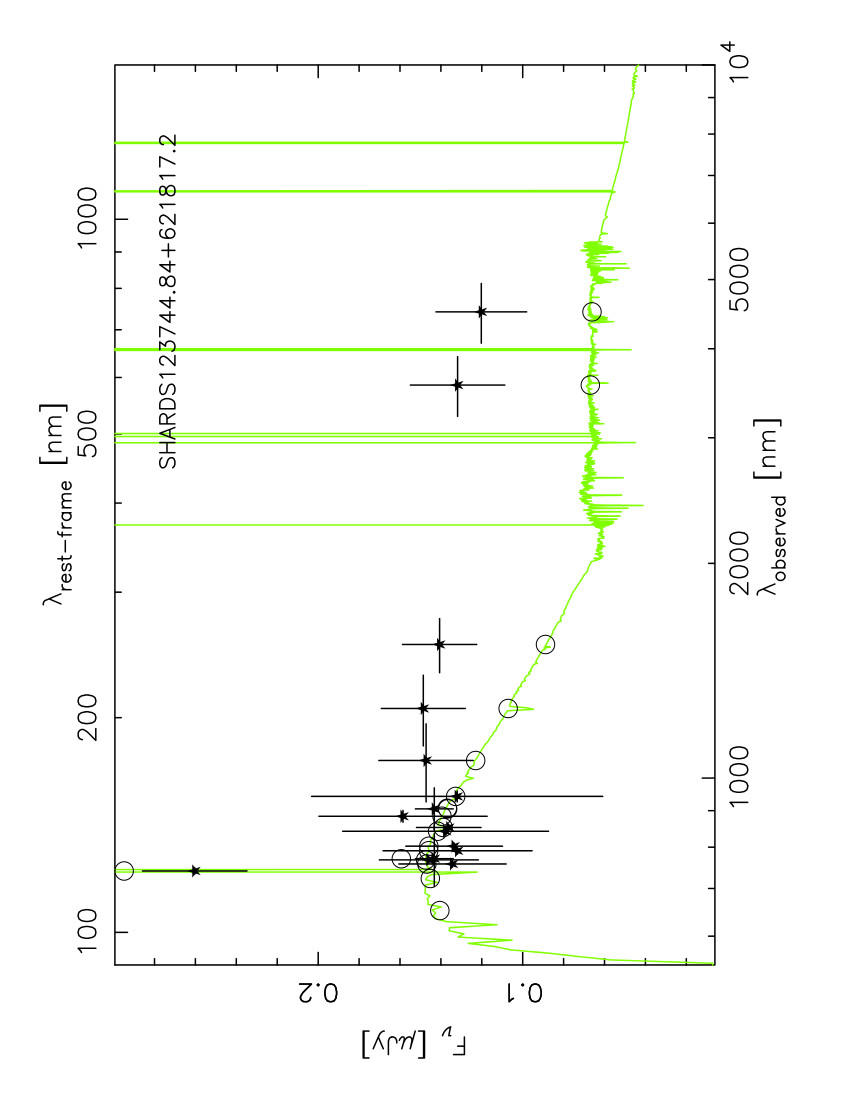}	
	\caption{ Model fit with a SSP for Obj2. Note the size of the deviation from a good fit starting at the optical-UV part of the spectrum, and becoming worse towards the rest-frame optical part of the spectrum. The contribution of nebular emission lines or the nebular continuum is negligible. A similar plot for Obj1 (not shown) leads to similar conclusions.}
	\label{SSP}
	\end{centering}
\end{figure}

\begin{figure}
	\begin{centering}
	\includegraphics[angle=0, width=\columnwidth]{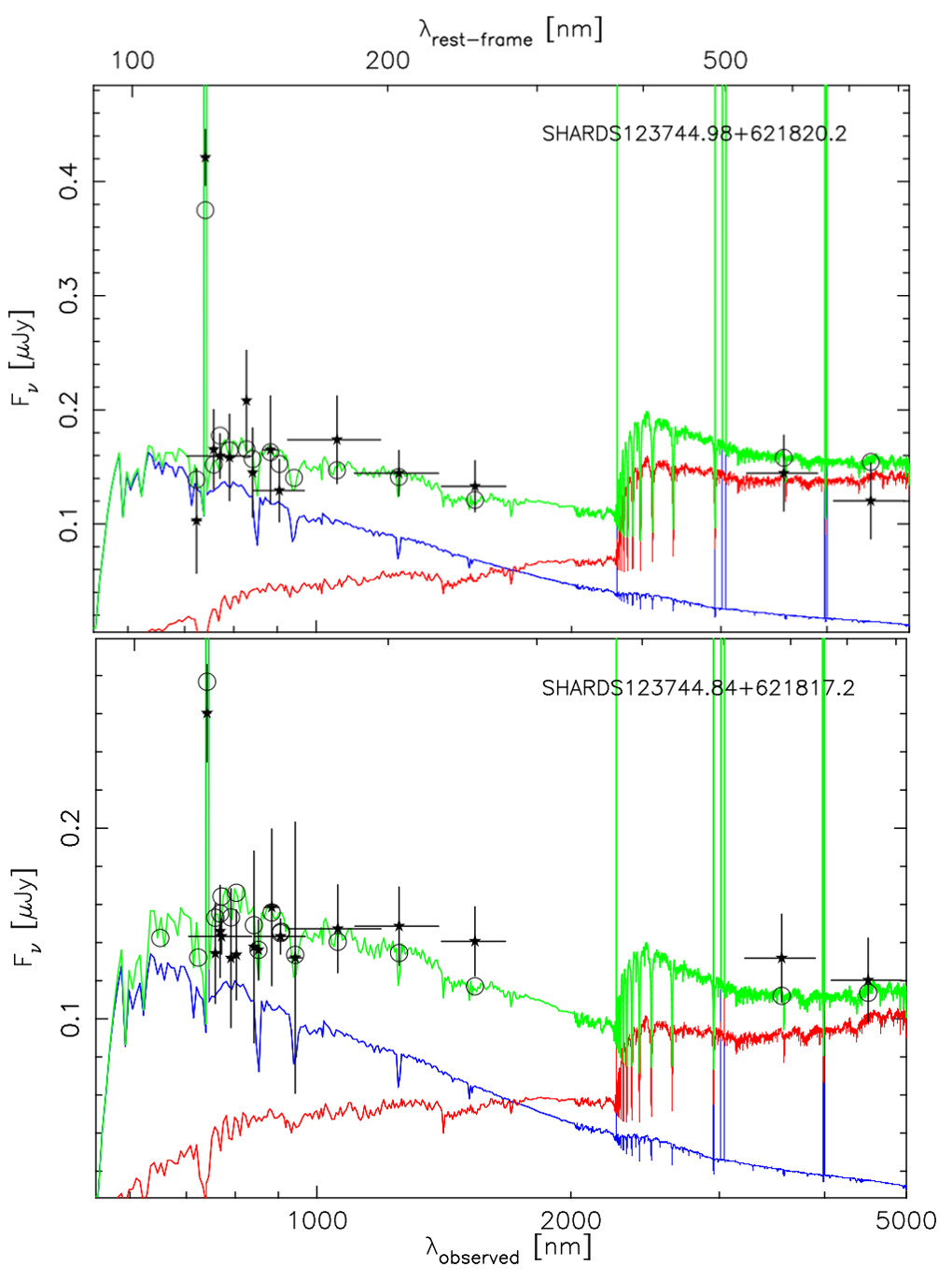} 
	\caption{Spectral energy distributions including all the data available for Obj1 (up) and Obj2 (down). Filled stars show the observed data, while open circles show the fluxes derived from the best-fitting stellar population model. The spectro-photometric SEDs have been fitted to a two-component stellar population model. Blue shows the contribution of the youngest component, in red the oldest, and in green the resulting fit.} 	
	\label{FullSEDs}
	\end{centering}	
\end{figure}

\subsection {Episodic star formation bursts}
SEDs of the two SHARDS sources have been measured, including all the data available in Rainbow\footnote[2]{The Rainbow Cosmological Surveys Database is operated by the Universidad Complutense de Madrid (UCM), partnered with the University of California Observatories at Santa Cruz (UCO/Lick,UCSC)}, namely SHARDS, {\it HST (ACS \& WFC3)} and {\it Spitzer IRAC} photometry. We have fitted the photometry together with the spectroscopically measured Ly$\rm{\alpha}$ fluxes to models based on Single Stellar Populations (SSPs). {\bf Our modelling procedure starts from the \citet{Bruzual&Charlot2003} stellar emission library and adds nebular continuum emission as well as hydrogen and helium emission lines, as described in \citet{Perez-Gonzalez2003}. Briefly, from the number of Lyman photons predicted from the stellar population models we calculate the gas emission using \citet{Ferland1980}. Emission lines use the relations in Brocklehurst (1971) and the theoretical line ratios expected for a low-density gas (n$_e$=10$^2$~cm$^{-3}$) with T$_e$=10$^4$~K (case B recombination). No metal lines have been modelled, as the calculations would involve further assumptions about metallicities and other uncertain parameters.} {\bf We assumed a delayed exponentially declining Star Formation History (SFH), parameterized by the decay time $\rm{\tau}$, and have run models varying the decaying factor, age and metallicity. The models use a \citet{Chabrier2003} IMF and the \citet{Calzetti2000} extinction law.} The models also assume 15\% escape of ionizing photons, hence only 85\%  interact with the ISM and contribute to the production of Ly$\rm{\alpha}$ photons that are then allowed to escape without restriction (fesc(Ly$\rm{\alpha}) =1$). {\bf We tried initially to fit the data with a single star formation episode extended in time (Fig.~\ref{SSP}), but we could not fit simultaneously the Ly$\rm{\alpha}$ emission and the rest-frame optical (IRAC) data, underestimating the optical rest-frame emission by about a factor of three. The contribution of the nebular emission is negligible since the redshifted H$\rm{\alpha}$ line falls in between the IRAC (3.6$\rm{\mu}$m) and (4.5$\rm{\mu}$m) filters, and the estimated H \& He nebular continuum would at most add 20\% emission to these filters \citep{Zackrisson2008}.} {\bf This discrepancy led us to explore models with two stellar populations in different evolutionary states. Since the parameter space is rather degenerate we decided to constrain at least the extinction to be mild, as expected at z$\sim$~5. We have then run models letting all parameters (age of the burst, decay time, metallicity, and mass) free for each population component, and one more parameter to compare the stellar masses of the two populations. With these constraints, the fitting procedure converges for both objects to a very recent ($\sim$~2~Myr) burst of star formation, coexisting with an underlying older population formed some 100~Myr ago (Fig.~\ref{FullSEDs}). The young stellar populations account well for the Ly$\rm{\alpha}$ emission in each galaxy, while the underlying populations fit well the rest-frame IRAC continuum. These underlying older stellar populations formed around 127$_{-14}^{+17}$ and 66$_{-26}^{+32}$~Myr for Obj1 and Obj2 respectively, and are now unable to keep ionizing the ISM. They are nonetheless responsible for the bulk of the mass of each of these galaxies.}

{\bf The SED fitting procedure assumes that no Ly$\rm{\alpha}$ photons are destroyed by interactions with the neutral medium. We have performed a second iteration using the Ly$\rm{\alpha}$ equivalent widths to further constrain both the Ly$\rm{\alpha}$ escape fraction and the evolutionary state of the SSPs in each galaxy.  \citet{Pena-Guerrero&Leitherer2013} have recently published improved predictions of EW(Ly$\rm{\alpha}$) values, based on a library of Ly$\rm{\alpha}$ stellar profiles combined with the nebular emission originated by the ionizing flux, for different values of the Ly$\rm{\alpha}$ escape fraction. Using the measured values (Tab.~\ref{Tab2}), and assuming   very young populations ($\sim$2~Myr), the  Ly$\rm{\alpha}$ escape fractions would be constrained to upper limits of $\sim$~0.25 (Obj1) and $\sim$~0.15 (Obj2)}. Note that \citet{Hayes2011} estimated an average fesc(Ly$\rm{\alpha}$) value of 0.2 at z=5, which is perfectly consistent with these results. Tab.~\ref{Tab3} summarizes the best fitting model parameters for both galaxies and Figure ~\ref{FullSEDs} shows the resulting best fits. Obj1 shows stronger Ly$\rm{\alpha}$ emission (both in luminosity and equivalent width) than Obj2, though both starburst episodes are very similar. This difference is likely due to the presence of some mild obscuration in Obj2, which leads to lower values of the Ly$\rm{\alpha}$ escape fraction.  
 
\begin{table}
\caption{Parameters of the resulting best fitting stellar populations. Note that the ${\tau}$ value for the old population of Obj2 is rather uncertain. We are witnessing a very young burst of star formation. At this age it is difficult to say whether the burst will decay fast or will last a long time. The ${\tau}$ derived is indeed consistent with a constant SFH}

\label{Tab3}
\begin{center}
\begin{tabular}{@{}ccccc}
 \hline
 &\multicolumn{2}{c}{Obj1} & \multicolumn{2}{c}{Obj2} \\ 
  & young & old & young & old \\ \hline
 Z(Z$_{\sun}$) & 1.0 & 1.0 &  1.0 & 1.0 \\
 \vspace{2pt}
  $\tau$(Myr) & $\rm{1.2_{-0.2}^{+0.3}}$ & $4.0_{-0.8}^{+1.1} $ & ${11_{-9}^{+50}}$ & 5.9$_{-1.8}^{+6.1}$ \\
 \vspace{2pt}
  Age (Myr) & 1.7$_{-0.2}^{+0.3}$ & 127$_{-14}^{+17}$ &  1.6$_{-0.5}^{+0.7}$ & 66$_{-26}^{+32}$ \\
  \vspace{2pt}
    A$_v$ (mag) & 0.0$_{-0.0}^{+0.1}$ & 0.0$_{-0.0}^{+0.1}$ &  0.1$_{-0.1}^{+0.1}$ & 0.1$_{-0.1}^{+0.2}$ \\
  \vspace{2pt}
  M (log M$_{\sun}$) & 7.8$_{-0.2}^{+0.2}$ & 9.3$_{-0.1}^{+0.1}$ &  7.8$_{-0.3}^{+0.3}$ & 8.9$_{-0.2}^{+0.1}$ \\
 \end{tabular}
 \end{center}
 \end{table}
  
Note, as mentioned before, that the solutions presented are rather degenerate, though they are the best solutions we obtain with the available data. With this caveat, we claim to have found two examples of LBGs with Ly$\rm{\alpha}$ emission who have experienced each two distinct episodes of star formation, the current one and a former one held some 100~Myr ago.  \citet[][and references therein]{Stark2009} show that episodic star formation explains better the evolution of the $M_{\star}-M_{1500}$ relation, the UV LF and the clustering properties of z$<4$ sources, while constant SF modes fail to reproduce the observed stellar mass growth between z $\simeq 5$ and 6. Here, we show two cases of galaxies undergoing episodic star formation. Indeed, having spectroscopy that reveals the Ly$\rm{\alpha}$ emission, and therefore the young stellar population, in combination with broad band photometry up to 750~nm (rest-frame) that describes the older populations, allows a much better determination of the SFH of these two galaxies. It is still pending however to test whether this is more common in high z galaxies or it is rather due to the fact that, in this particular case, these galaxies seem to be involved in a threefold interaction, which includes a possible merger between two of them. 

\section{Summary}

In a search for Ly$\rm{\alpha}$ emitting sources in the SHARDS survey, two sources were spotted in filter SF738W17, which were not present in any bluer filters. We have studied these two sources spectroscopically, confirming their similar redshifts at z=5.07, and have measured their Ly$\rm{\alpha}$ emission, which allows constraining the evolutionary state of the ionizing stellar population. The combination of the spectroscopic data with multi-band photometry from the rest frame UV up to the optical, reveals the presence of two distinct stellar populations in each galaxy. A very young population responsible for the Ly$\rm{\alpha}$ emission and the UV continuum with very small masses involved, plus an older population, formed $\sim$100~Myr before, that accounts for both the rest-frame optical emission and the bulk of the masses of these galaxies. These are therefore two examples of galaxies undergoing episodic star formation. Besides, the two sources are very close both in the sky and velocity, and one of the sources is in turn interacting/merging with yet another source. Whether episodic star formation is  common in high-z galaxies is however still to be seen, as in our case the sources seem to be undergoing an interaction-merger, that may have been responsible for the recent onset of star formation.

\section*{Acknowledgments}
JMRE, OGM \& NCR (CDS2006-00070, AYA2012-39168-C03-01), CMT and JMMH (AYA2010-21887-C04-04, AYA2010-21887-C04-02 and AYA2012-39362-C02-01) and AHC (AYA2012-31447) acknowledge support from the Spanish MINECO under the PNAyA grants given in parenthesis. OGM acknowledges a Juan de La Cierva Fellowship (MINECO). We are very grateful to the GTC support staff for their help with the observations. SHARDS is funded by the Spanish MINECO with grant AYA2012-31277. The Rainbow database is partially funded by the Spanish MINECO under grant AYA2012-31277 and the US NSF grant AST-08-08133. Based on observations made with the Gran Telescopio Canarias (GTC), installed at the Spanish Observatorio del Roque de los Muchachos of the Instituto de Astrof\'isica de Canarias, in the island of La Palma.

\label{lastpage}

\end{document}